\documentclass[preprint,prb,aps,superscriptaddress,showpacs,citeautoscript]{revtex4-1}
\usepackage{graphicx}
\usepackage{amsmath}
\usepackage{amssymb}
\usepackage{color}
\usepackage{dcolumn}
\usepackage{epsfig}
\usepackage{bm}
\usepackage{float}
\usepackage{silence}
\begin{document}

\title{Formation of As-As Interlayer Bonding  in the collapsed tetragonal phase of NaFe$_{2}$As$_{2}$ under pressure}

\author{Elissaios Stavrou}
\affiliation{Geophysical Laboratory, Carnegie Institution of Washington, Washington, D.C. 20015, U.S.A.}

\author {Xiao-Jia Chen}
\affiliation{Center for High Pressure Science and Technology Advanced Research, Shanghai 201203, China}
\affiliation{Geophysical Laboratory, Carnegie Institution of Washington, Washington, D.C. 20015, U.S.A.}
\affiliation{Key Laboratory of Materials Physics, Institute of Solid State Physics, CAS, Hefei, 230031, China}

\author {Artem R. Oganov}
\affiliation{Department of Geosciences, Center for Materials by Design, Institute for Advanced Computational Science, Stony Brook University, Stony Brook, NY 11794-2100, U.S.A.}
\affiliation{Moscow Institute of Physics and Technology, 9 Institutskiy lane, Dolgoprudny city, Moscow Region, 141700, Russian Federation}
\affiliation{School of Materials Science, Northwestern Polytechnical University, Xi'an,710072, China}

\author{A. F. Wang}
\affiliation {Hefei National Laboratory for Physical Science at Microscale and Department of Physics, University of Science and Technology of China, Hefei, Anhui 230026, China}

\author {Y.  Yan}
\affiliation {Hefei National Laboratory for Physical Science at Microscale and Department of Physics, University of Science and Technology of China, Hefei, Anhui 230026, China}
\author {X. G. Luo}
\affiliation {Hefei National Laboratory for Physical Science at Microscale and Department of Physics, University of Science and Technology of China, Hefei, Anhui 230026, China}

\author{X. H. Chen}
\affiliation {Hefei National Laboratory for Physical Science at Microscale and Department of Physics, University of Science and Technology of China, Hefei, Anhui 230026, China}

\author {Alexander F. Goncharov}
\affiliation{Geophysical Laboratory, Carnegie Institution of Washington, Washington, D.C. 20015, U.S.A.}
\affiliation{Key Laboratory of Materials Physics, Institute of Solid State Physics, CAS, Hefei, 230031, China}
\affiliation{University of Science and Technology of China, Hefei, 230026, China}

\begin{abstract}
NaFe$_{2}$As$_{2}$ is investigated experimentally using powder x-ray diffraction and Raman spectroscopy at pressures up to 23 GPa  at room temperature and using ab-initio calculations.  The results reveal a pressure-induced structural modification at 4 GPa from the starting tetragonal to a collapsed tetragonal phase. We determined the changes in interatomic distances under pressure that allowed us to connect the structural changes and superconductivity. The transition is related to the formation of interlayer As-As bonds at the expense of weakening of Fe-As bonds in agreement with recent theoretical predictions.
\end{abstract}

\maketitle

\section{Introduction}
Nature of superconductivity in iron-based superconductors (FeSCs) remain a challenge of  modern condensed matter physics due to the complicated interplay between their structure, magnetism, electronic nematicity, and superconductivity \cite{Mazin2010, Paglione2010}. A generic picture for the early discovered sister cuprate system is that superconductivity emerges and develops in a dome in the charge carrier concentration versus temperature phase diagram when magnetism of a parent compound is suppressed and a narrow spin glass state \cite{Dioguardi2013} is passed. To study the emergence of superconductivity, one can naturally apply chemical doping, however application of pressure has the similar effect in  Fe SCs as there are some similarities for the structural distortions between pressure and chemical doping \cite{Kimber2009}. In the case of 122 Fe SCs, pressure-induced suppression of antiferromagnetic spin fluctuations was proposed to account for the disappearance of superconductivity in a so-called collapsed tetragonal (CT) phase \cite{Pratt2009}. However, the  structural modification has been suggested to be a key to the understanding of the superconductivity of iron pnictides \cite{Stewart2011}. A common structural characteristic of all Fe SCs is the presence of As-Fe-As layers forming edge-sharing FeAs$_4$ tetrahedra. It is believed that superconductivity emerges from these layers \cite{Mazin2010, Paglione2010}.

NaFe$_{2}$As$_{2}$ crystallizes in the tetragonal (T) ThCr$_2$Si$_2$-type structure (\emph{I4/mmm}) \cite{Friederichs2012,Gooch2010} (see Fig. 1), belonging to the so-called 122 family of FeSCs . It has been characterized as the  \textquotedblleft missing\textquotedblright member of the 122 system, since initially it was thought that Na, due to its small ionic radius,  cannot fill the coordination sphere \cite{Friederichs2012,Gooch2010} between the As-Fe-As layers. Moreover, NaFe$_{2}$As$_{2}$ has a  remarkably high superconducting transition temperature $T_c$ of 25 K \cite{Gooch2010} in comparison with less than  5 K in the case of Cs and Rb within this AFe$_2$As$_2$ family.

\begin{figure}[ht]
{\includegraphics[width=100mm]{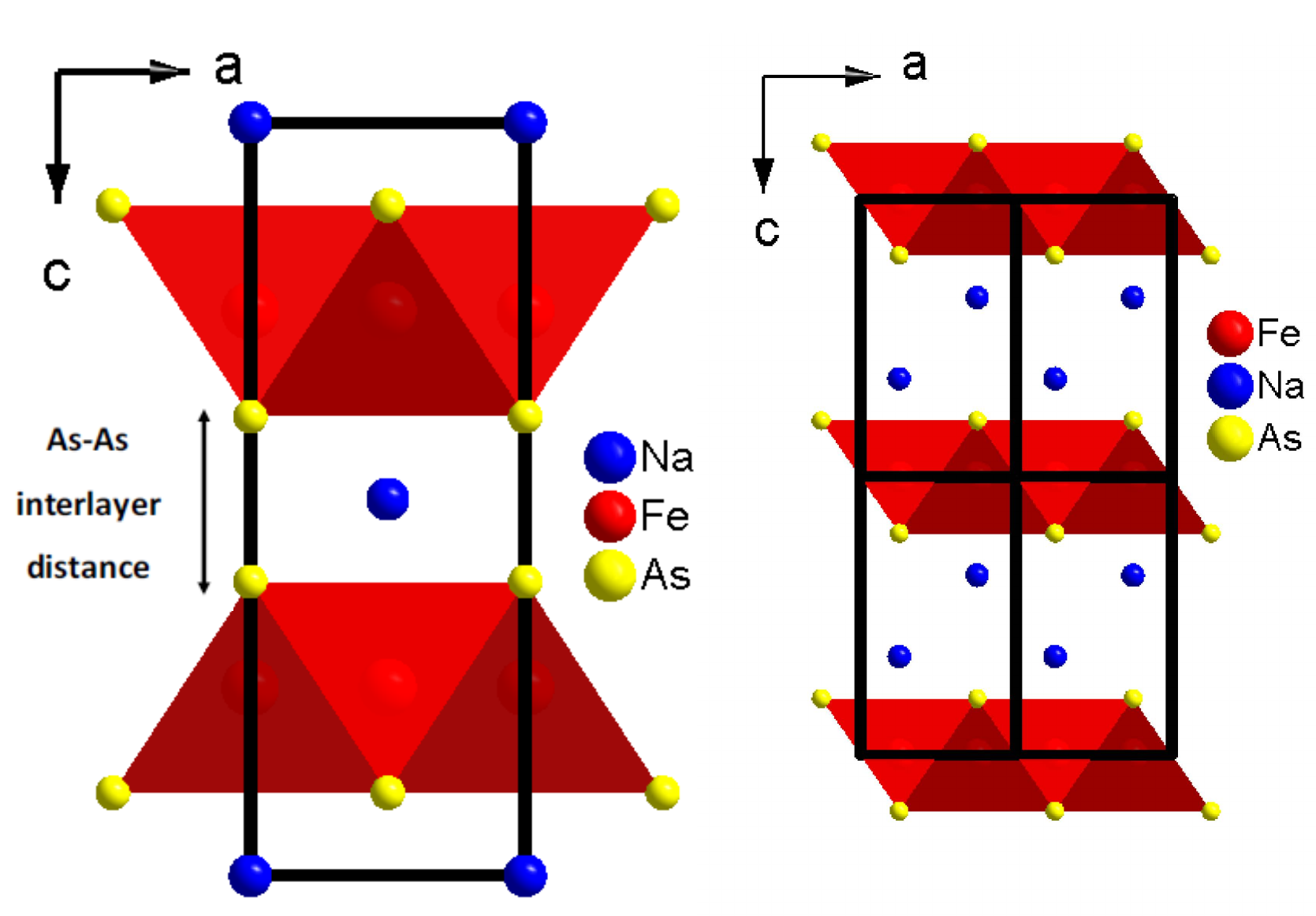}}
\caption{(a) ThCr$_2$Si$_2$-type crystal structure of NaFe$_2$As$_2$, (b) PbClF-type crystal structure of NaFeAs. }
\end{figure}

Previously reported structural studies on 122 FeSCs under pressure (e.g. Ref \cite{Mittal2011}) were limited to the cases of A being a divalent  alkaline earth or rare earth metal atom, while systems with monovalent alkali metal, to the best of our knowledge, have not been studied. Although  all previous studies suggested that the transformation from the T to the  CT phase  is a universal characteristic of divalent AFe$_{2}$As$_{2}$ compounds \cite{Uhoya2010}, a detailed structural study on the atomic level and the link between structural modifications and superconductivity under pressure is still missing.  In this paper we present a combined experimental, using x-ray diffraction (XRD) and Raman spectroscopy, and theoretical study of  NaFe$_{2}$As$_{2}$ under pressure. We examine the structural behavior of AFe$_2$As$_2$ superconductors  under pressure in the case of A being an alkali metal  to obtain a deeper insight on the interplay between structure and superconductivity under pressure. We present, for the first time, direct experimental evidence of the formation of As-As interlayer bonds under pressure. According to theoretical predictions \cite{Colonna2011, Yildirim2009} this is the key parameter determining the correlation between structure and superconductivity for 122 SC.

\section{Experimental and computational methods}

Fine powder prepared of NaFe$_{2}$As$_{2}$  single crystals were  loaded into a diamond anvil cell (DAC) for the angle dispersive XRD experiments. For the high-pressure Raman measurements, small  pieces with a typical dimension of 50 $\mu$m were loaded into a DAC. Neon was used as the pressure-transmitting medium for both sets of measurements. Small quantities of ruby and gold powder were also loaded, for determination of pressure through ruby luminescence and gold equation of state (EOS), respectively. XRD data were collected at the GSECARS (sector 13), using a MAR355 CCD detector. The monochromatic x-ray beam ($\lambda$=0.3344 \AA) was focused to a nominal diameter of 4 $\mu$m. The images were integrated using the FIT2D \cite{Hammersley1996} program to yield intensity versus 2$\theta$ diagrams.
Raman spectra were measured in a 135 degree \cite{Goncharov2003}  geometry using the 532 nm line from a solid state laser for excitation. Ultralow fluorescence type IIa synthetic diamond anvils were used for Raman experiments.

Single crystals in the Na-Fe-As system were grown by use of the NaAs flux method. We obtained NaAs by reacting the mixture of the elemental Na and As in an evacuated quartz tube at 200 $^o$C for 10 h. NaAs and Fe powders were carefully weighed  and thoroughly ground. The mixtures were put into alumina crucibles and then sealed in iron crucibles under 1.5 atm of highly pure argon gas. The crucibles were sealed and heated to 950 $^o$C at a rate of 60 $^o$C/h in the tube furnace filled with the inert atmosphere and kept at 950 $^o$C for 10 h and then cooled slowly to 600 $^o$C at 3 $^o$C/h to grow single crystals. The shiny crystals with typical size of 6$\times$6$\times$0.2 mm can be easily cleaved from the melt.

Our theoretical calculations were performed using density functional theory within the generalized gradient approximation \cite{Perdew1996} and using the PAW method \cite{kresse1999} as implemented in the VASP code \cite{Kresse1996}. The PAW potentials used had [Be] core (outermost radius 2.2 a.u.) for Na atoms, [Ar] core (radius 2.3 a.u.) for Fe atoms, and [Ni] core (radius 2.1 a.u.) for As atoms. We used the plane wave kinetic energy cutoff of 360 eV and 6$\times$6$\times$3 $\Gamma$-centered meshes for Brillouin zone sampling. Electronic optimization was done self-consistently with a threshold of 10$^{-4}$ eV/cell, and structure relaxation proceeded until changes in the enthalpy were below 10$^{-3}$ eV/cell.  A ferromagnetic configuration with all Fe atoms given a starting magnetic moment of 2 Bohr magnetons was used. The starting magnetic moment plays little role (only the magnetic symmetry is important) and was optimized during electronic and structural relaxation.

\section{Results}

\textbf{X-ray diffraction results and analysis}

Figure 2 shows selected XRD patterns of NaFe$_{2}$As$_{2}$ obtained at various pressures up to 23 GPa. All the observed peaks in this pressure range can be properly indexed according to the tetragonal ThCr$_2$Si$_2$-type structure. On the other hand, an unusual pressure behavior of the Bragg peaks related with the $a$-axis can be clearly seen at low pressures. This is highlighted by the merge of the (114) and (200) peaks  at 2$\theta$ $\approx10^o$ and the splitting of the peak at  2$\theta$ $\approx7^o$ into its components, (110) and (103). We observe a decrease of the 2$\theta$ position  of the (200) peak and a practically constant position of the (101) peak, while the 2$\theta$ positions of  all other peaks [\emph{e.g.} (002)] shift towards higher values.

\begin{figure}[ht]
{\includegraphics[width=100mm]{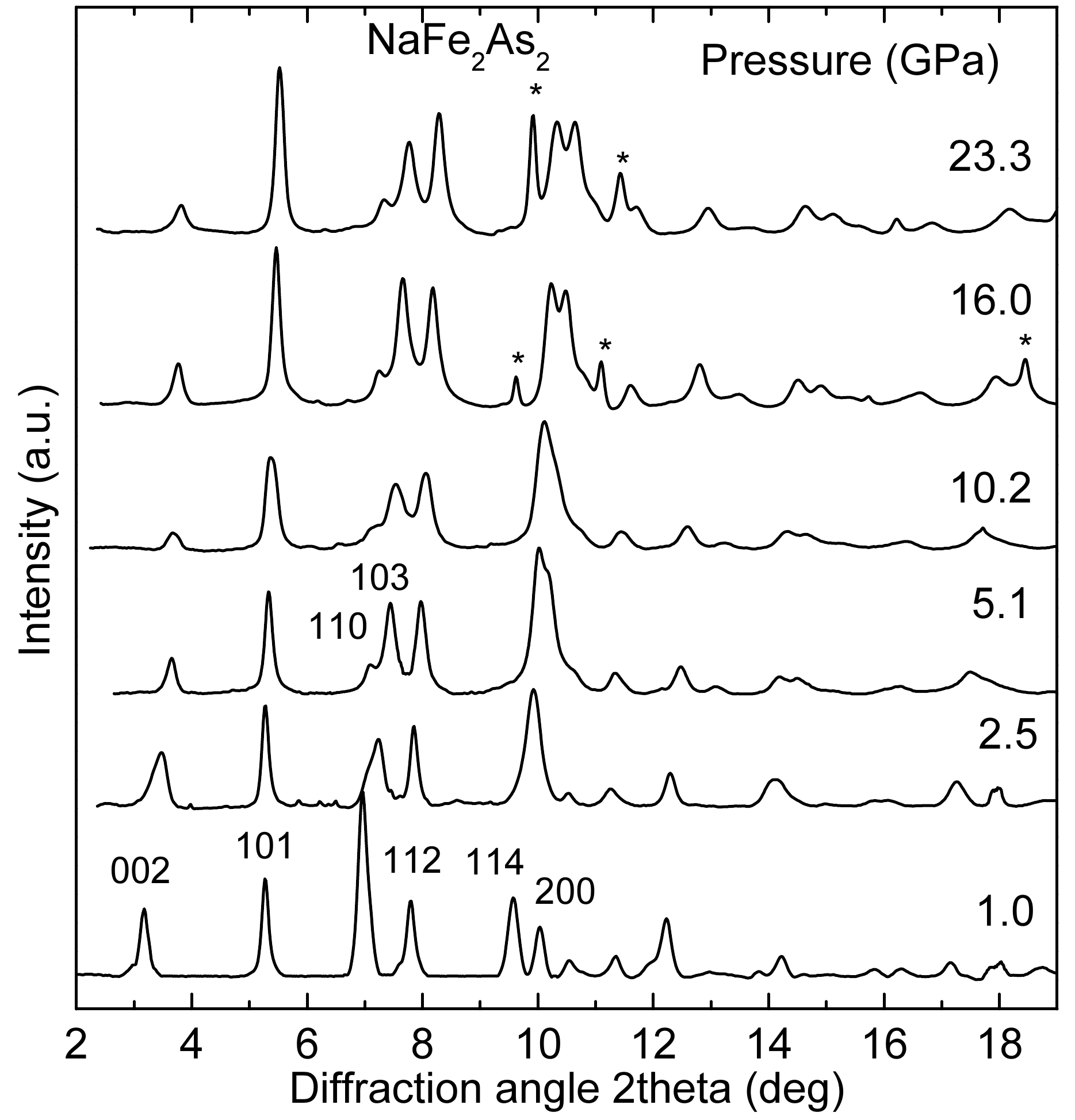}}
\caption{X-ray diffraction patterns of NaFe$_{2}$As$_{2}$ at various pressures. The asterisks denote peaks from Ne. }
\end{figure}

In order to determine the positional parameters and the interatomic distances we have performed a detailed Rietveld refinement of the XRD patterns
using the GSAS program  \cite{Larson2004}. Typical refined profiles are shown in Fig. 3 at 1.0 and 5.1 GPa.
The only free positional parameter of this structure is the $z$ position of As atoms, Wyckoff position (WP) 4$e$ (0, 0, $z$), since positions of Na and Fe atoms are fixed. Our two-dimensional (2D) images reveal an almost perfect ring shape of reflections, mainly due to the very fine powder used in this study. In addition wide angle dispersive XRD measurements helped us to overcome any intensity redistribution caused by preferred orientation. Moreover, XRD patterns have been collected from several spots of the sample at the same pressure, and the results were averaged.
The agreement between experimental and theoretical interatomic distances provides an additional confidence of these results.

\begin{figure}[ht]
{\includegraphics[width=100mm]{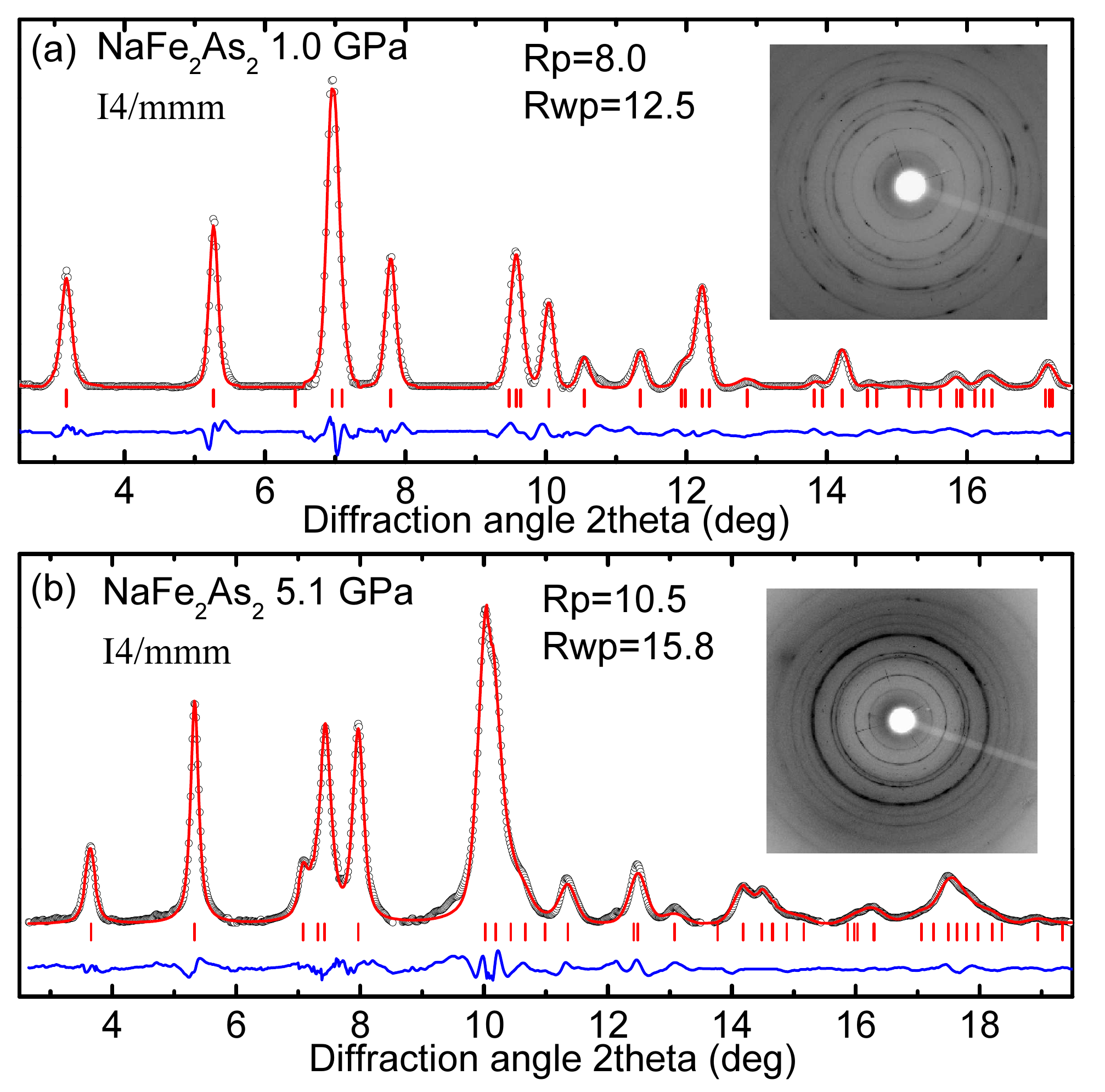}}
\caption{Rietveld refinements for NaFe$_{2}$As$_{2}$ at (a) 1.0 GPa and (b) 5.1 GPa. Respective 2D x-ray diffraction images are shown in insets. }
\end{figure}

From XRD data of  NaFe$_{2}$As$_{2}$  we have obtained the lattice parameters and the unit cell volume as a function of pressure and the results are shown in the plots of Fig. 4. The pressure evolution of the lattice parameters and the $c/a$ ratio changes at  a critical pressure $P_c$ $\approx$ 4 GPa. Below $P_c$, $a$-axis increases while both the $c$-axis and the $c/a$ ratio decrease rapidly.  In contrast above $P_c$, $a$-axis starts to decrease and $c$-axis becomes much less compressible. Consequently the $c/a$ ratio shows a gradual decrease.  The different behavior of $c$-axis, below and above $P_c$, can be viewed as a modification of the initial tetragonal phase (T) to the so-called collapsed tetragonal (CT) phase (\emph{e.g.} Ref. \cite{Kreyssig2008}). We have fitted the pressure-volume data by the third-order Birch equation of state \cite{Birch1978} and determined the bulk modulus $B_0$ and its first pressure derivative $B_0'$ (Fig. 4). The CT phase is much less compressible compared to the T phase as  $B_0$ increases by a factor of 5. No apparent volume discontinuity is observed at this isostructural phase transition.

\begin{figure}[ht]
{\includegraphics[width=100mm]{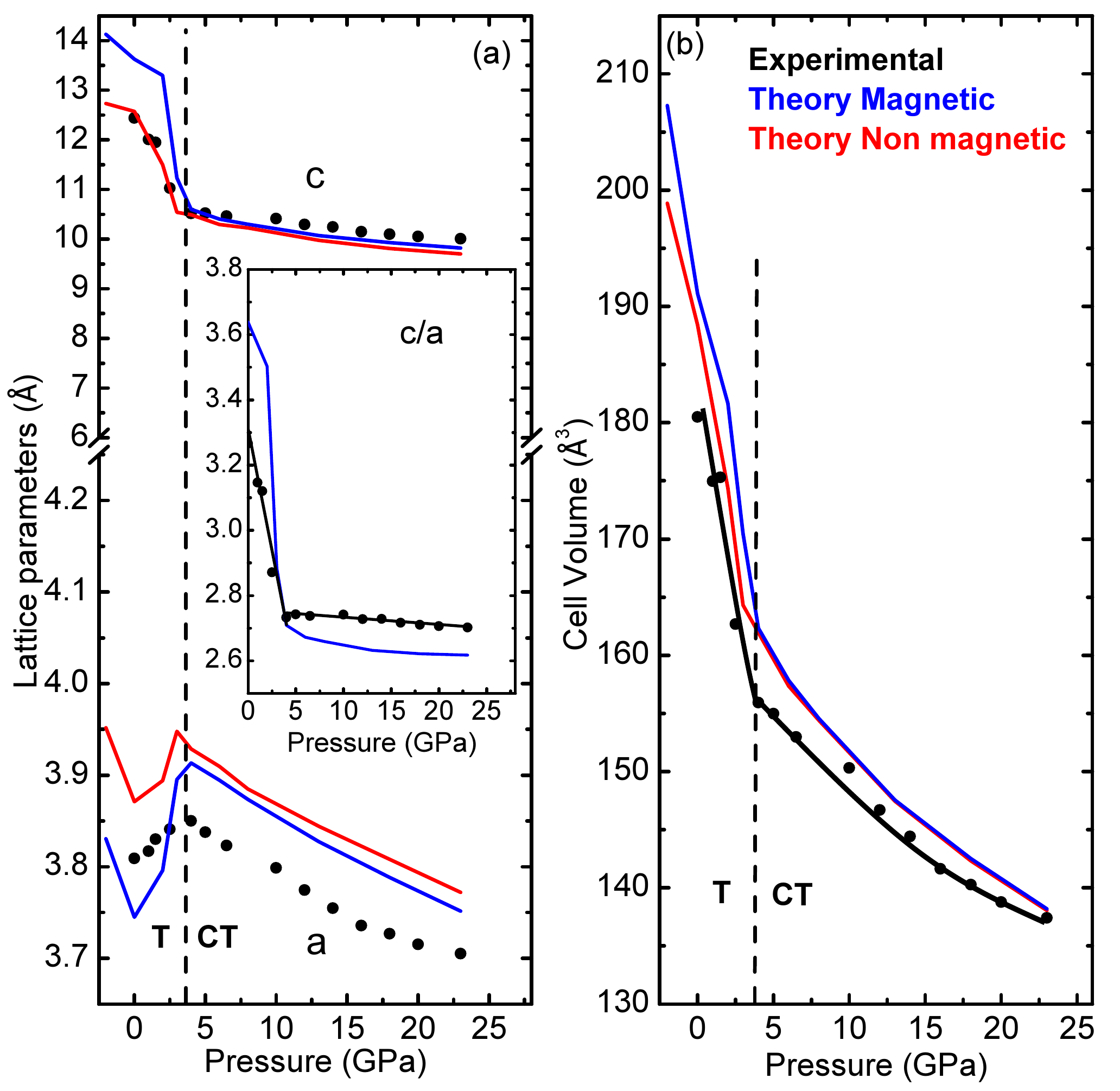}}
\caption{Pressure dependence of (a) lattice parameters and (b) cell volume of NaFe$_{2}$As$_{2}$. The inset shows the pressure dependence of the $c$/$a$ ratio. Experimental data are shown with solid symbols and theoretical predictions (shifted by 4 GPa to higher P, see text) with blue (M-phase) and red (NM-phase)lines, respectively. The solid black curves in (b) are the third-order Birch-Murnagan EOS fits for the T ($B_0$=22 GPa, $B'$= 4) and CT ($B_0$=110 GPa, $B'$= 4.5) phases.}
\end{figure}

The results from the theoretical predictions  are also plotted in Fig. 4 for both a magnetic (M) and a non-magnetic phase (NM). In the case of the M phase, a ferromagnetic configuration has been used and the energy difference from the antiferromagnetic configuration is very small. It is well known that density-functional calculations have a systematic error of several GPa in the equation of state, and a simple constant shift usually brings computed equations of state into close agreement with experiment \cite{Oganov2001}. Here, the theoretical pressures were corrected by 4 GPa. With this correction, an almost perfect agreement between experimental and theoretical values (for both phases) can be clearly seen from Fig. 4. It is noteworthy that the M phase,  although it has higher volume at low pressures, is more stable than the NM one, due to the magnetic interactions which result in a (slightly) lower enthalpy up to well above $P_c$. Moreover, both phases exhibit a T to CT transformation at the same pressure. In the magnetic case (which corresponds to observations of Ref. \cite{Friederichs2012}) our calculations suggest that CT phase remains ferromagnetic at T = 0 K. Since theory predicts the same phase transformation in magnetic and nonmagnetic cases, we can conclude that the structural behavior is not driven by magnetic interactions.

\textbf{Raman spectroscopy}

Raman spectra of  NaFe$_{2}$As$_{2}$  under pressure show three out of four Raman-active zone-center modes predicted from group theory \cite{Litvinchuk2008}: $A_{1g}$ (As(z)), $B_{1g}$(Fe(z))and $E_g$(As-Fe(xy)and Fe-As(xy)) (Fig. 5(a)). $A_{1g}$ and $B_{1g}$ modes correspond to the displacement of As and Fe atoms along the $c$-axis, respectively. The two $E_g$ modes correspond to mixed lateral As and Fe displacements inside the As and Fe layers.
The ambient pressure frequencies are in very good agreement with previous studies \cite{Litvinchuk2008} with the exception of the very low intensity high frequency $E_g$(Fe-As(xy)) mode, which was not observed. All Raman modes show normal mode behavior \emph{i.e.} increase of the frequency with increasing pressure in phase T (Fig. 5(b)). At 3 GPa, a critical pressure which is in agreement with that determined from XRD measurements, there is an apparent change in the pressure slope of Raman modes (see Table I), accompanied by an  increase of the linewidth. Remarkably, the  $B_{1g}$(Fe) mode shows a softening in CT phase. The mode   Gr\"{u}neisen parameters ($\gamma_T$)  determined using the experimental results of this work  are shown Table I. The $\gamma_T$ parameters of modes of the T phase are common for materials  with mixed covalent-ionic bonding (such as within the As-Fe-As layers), due to the presence of weaker interlayer bonds, which experience a larger compression.  In the CT phase,  the $A_{1g}$ (As)  mode exhibit a higher $\gamma_T$ which almost doubles. These changes suggests an alteration of this mode character from intralayer to interlayer , which results in an increase of the contraction of the respective bond. Similarly, we find the doubling of the $\gamma_T$ of the  intralayer $E_g$ (As-Fe) mode.  It is the most intriguing that the $\gamma_T$ of the $B_{1g}$ (Fe) mode becomes negative, in CT phase suggesting weakening of  bonds under pressure.

\begin{figure}[ht]
{\includegraphics[width=100mm]{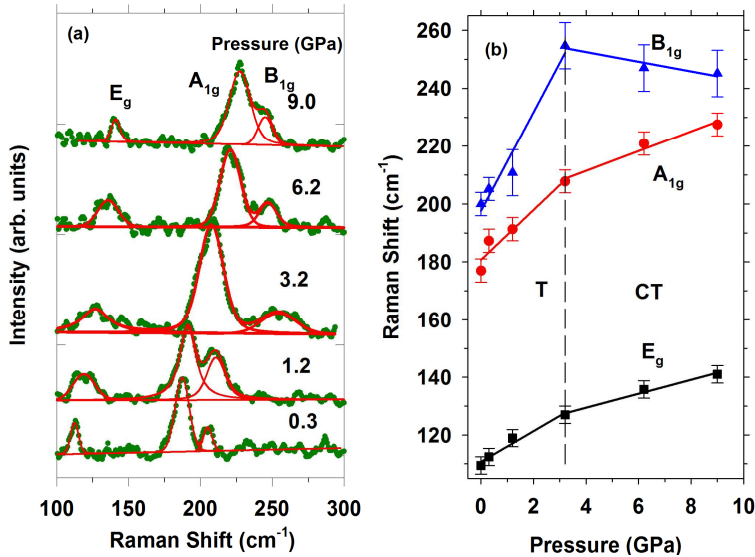}}
\caption{(a) Raman spectra of  NaFe$_{2}$As$_{2}$ at various pressures. (b) Raman peaks frequencies plotted against pressure. In (b) lines are the results of linear fits.  }
\end{figure}

\begin{table}[H]
  \caption{Assignments, ambient conditions frequencies $\omega_0$ (in cm$^{-1}$), slopes ($\partial\omega$/$\partial$$P$)$_T$ (in cm$^{-1}$GPa$^{-1}$) and Gr\"{u}neisen parameters  $\gamma$$_T$=-$\partial$(ln$\omega$)/$\partial$(ln$V$)$\mid_T$ of the Raman modes of the T and CT phases.}
  \centering
  \small
\begin{ruledtabular}
\begin{tabular}{l|lll|lll}
& & T && &CT	  &\\
\hline
Mode & $\omega_0$& ($\partial\omega$/$\partial$$P$)$_T$ & $\gamma$$_T$ & $\omega_0$& ($\partial\omega$/$\partial$$P$)$_T$& $\gamma$$_T$  \\
\hline
 $E_g$ & 110.7 &    \; \:  5.3 & 1.05 &	119.8 & \; \: 2.4 & 2.2 \\
 $A_{1g}$  & 180.7 & \; \: 8.6 &  1.05 & 197.8 & \; \: 3.4 & 1.9 \\
$B_{1g}$ & 200.9 & \; \: 17.0 & 1.86 & 259.2 & \; \: -1.7 & -0.72\\
\end{tabular}
\end{ruledtabular}
\end{table}

\section{Discussion}

Our x-ray diffraction results reveal the behavior of the As-As interlayer distance  in comparison with the Fe-As intratetrahedral distance (see  Fig. 6). Figure 7 shows the experimentally and theoretically determined values of these distances under pressure. An almost perfect agreement between the two sets of values is observed. The As-As interlayer distance decreases rapidly with pressure up to 4 GPa. This decrease is almost half of that of  c-axis and it accounts for the whole c-axis contraction while the thickness of As-layers (vertical distance between As-As layers of the same As-Fe-As block) remains  almost constant or even increases. Shortening of c-axis is driven exclusively  by the distance   between the tetrahedra blocks ($c$=2$d_{As-As(c)}$+2$d_{As-Fe-As (c)}$). The interlayer As-As and the intralayer Fe-As distances become practically equal above 4 GPa, \emph{i.e.} after the transformation to the CT phase. The As-As distance above 4 GPa takes a value (2.4 \AA \enspace in experiment and 2.6 \AA \enspace  in theory) which is consistent with the formation of As-As bond (covalent radius of As is 1.2 \AA\/ \cite{Cordero2008}). Thus, we conclude that a major change in bonding between  As atoms occurs under pressure. It can be viewed as a change from 4-fold to 5-fold coordination with formation of interlayer As-As bonds (see  Fig. 6(b)). This conclusion is further justified by the doubling of the $\gamma_T$ of the $A_{1g}$ mode after the phase transition. In contrast the decrease of the $\gamma_T$ of the $B_{1g}$ mode indicates weakening of the respective Fe bond.

\begin{figure}[ht]
{\includegraphics[width=100mm]{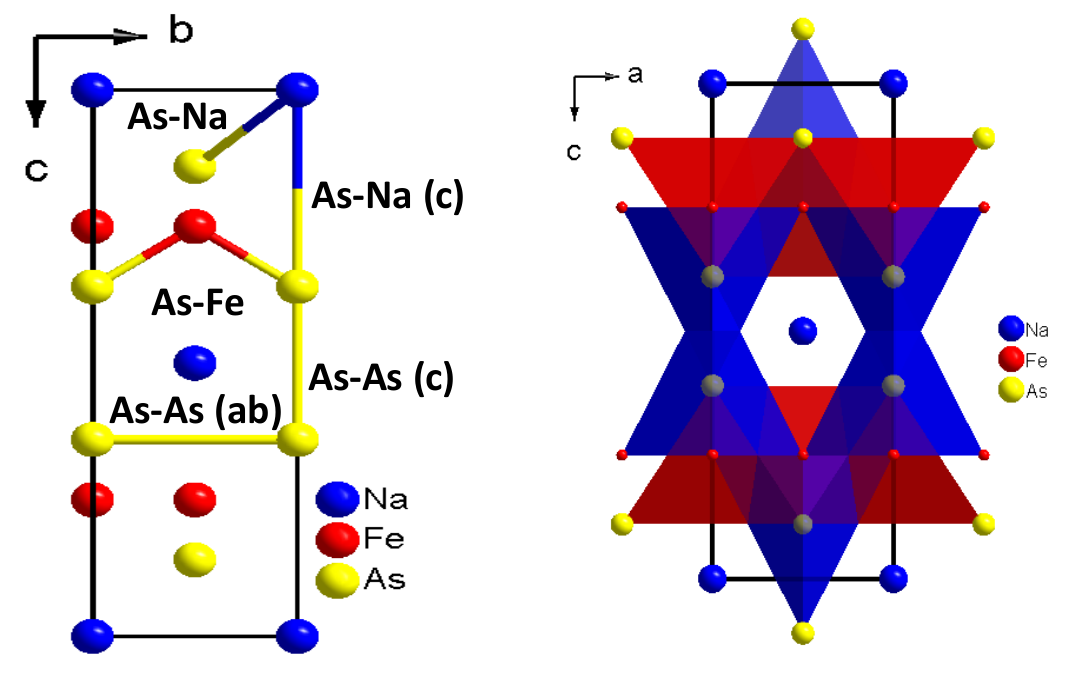}}
\caption{(a) ThCr$_2$Si$_2$-type crystal structure of NaFe$_2$As$_2$ showing the different interatomic distances and (b) ThCr$_2$Si$_2$-type crystal structure of NaFe$_2$As$_2$ with As atoms in 5-fold coordination. }
\end{figure}

\begin{figure}[ht]
{\includegraphics[width=100mm]{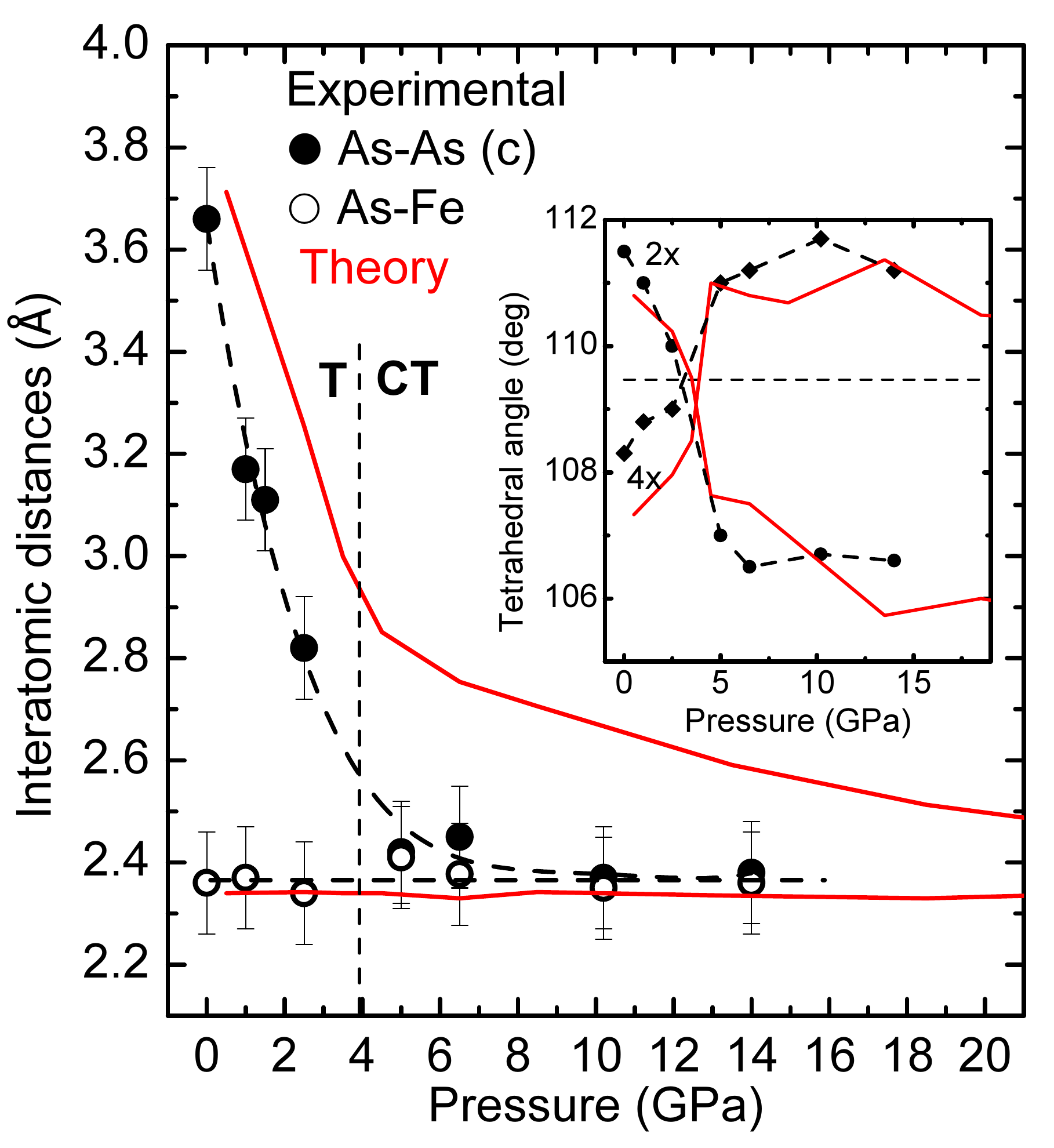}}
\caption{ Pressure dependence of selected interatomic distances of NaFe$_{2}$As$_{2}$: As-As interlayer along the $c$ axis and Fe-As inside the FeAs$_4$ tetrahedra.  Experimental and theoretical values are shown with black symbols and red lines respectively, the dashed lines are guides for the eye. The inset shows the pressure dependence of the two distinct tetrahedral As-Fe-As bond angles. The ideal tetrahedral angle is indicated by the horizontal dashed line. }
\end{figure}

Previous experimental studies on  AFe$_{2}$As$_{2}$ (A=Eu \cite{Uhoya2010}, Ba \cite{Uhoyab2010}, Sr \cite{Uhoya2011}, Ca \cite{Kreyssig2008}) compounds, reveal a common structural trend under pressure. This is an increase of $a$ while both the c and  $c/a$ ratio decrease rapidly  below $P_c$, and a decrease of $a$ and much less compressible $c$ above $P_c$. In the light of this observation,  Uhoya \emph{et al.} \cite{Uhoya2010} suggested that the transformation from the T to the  CT phase  is a universal characteristic of AFe$_{2}$As$_{2}$ compounds, while the value of $P_c$ depends on the divalent metal. Our study reveal a similar behavior for single valence metal suggesting that this universality is also independent of the valence of A metal. Transformation to the CT phase with further increasing pressure always suppresses \cite{Uhoyab2010} or diminishes superconductivity \cite{Uhoya2011}. In a study of CaFe${_2}$As${_2}$ under pressure \cite{Kreyssig2008} it has been suggested that one of the main structural differences between T and CT phases is the value of the As-Fe-As tetrahedral angles. Both tetrahedral angles change abruptly during the phase transition from a value very close  (T) to a value far away (CT) from the ideal tetrahedral angle (109.5$^o$). Since $T_c$ decreases with increasing deviation from the ideal  tetrahedral angles, it has been concluded that this is crucial for superconductivity although the direct link and the underling mechanism which correlates tetrahedral angles and superconductivity is not clear.

Recent theoretical studies \cite{Sanna2012,Colonna2011,Yildirim2009} were focused on the change in interatomic distances under pressure, mainly on the interlayer As-As distance(noted as As-As (c) in Fig. 6).
This distance has been predicted to decrease abruptly with pressure,  reaching a value that is very close to the As-As covalent bond distance in the CT phase. Consequently it has been proposed that the transition to the CT phase is induced by formation of a direct As-As interlayer bond and a weakening of the in-plane Fe-Fe  \cite{Colonna2011} and Fe-As bonds \cite{Yildirim2009}. It is noteworthy that in the case of 111 superconductors (e.g. Ref \cite{Zhou2012}) the Fe-As-Fe layers are separated by two layers of cations (see  Fig. 1). Thus, an interaction between As atoms of different layers is hindered and only intralayer modifications can affect superconductivity.

The spin-state of Fe is the key parameter which controls As-As bonding and consequently the lattice parameters \cite{Yildirim2009}. Under compression the magnetic moment  of Fe decreases and consequently, the strength of the Fe-As bonding (see Fig. 4 of Ref. \cite{Yildirim2009}). This is also evident from our calculations, see Fig. 8.  The results of our calculations suggest that the CT phase is magnetic $i.e.$ the Fe magnetic moment is not vanished but suppressed after the transformation decreasing the strength of the Fe-As bonding in agreement with softening of the $B_{1g}$ Raman mode (Fig. 5). The structural changes determined in this paper (Fig. 7) clearly support this scenario. A direct consequence of the As 5-fold coordination is that the system loses its two-dimensionality (Fig. 6(b)), and with it, superconductivity. Moreover the weakening of the in-plane Fe-Fe bonds \cite{Colonna2011} in the CT phase also decreases superconductivity.

 \begin{figure}[ht]
 {\includegraphics[width=100mm]{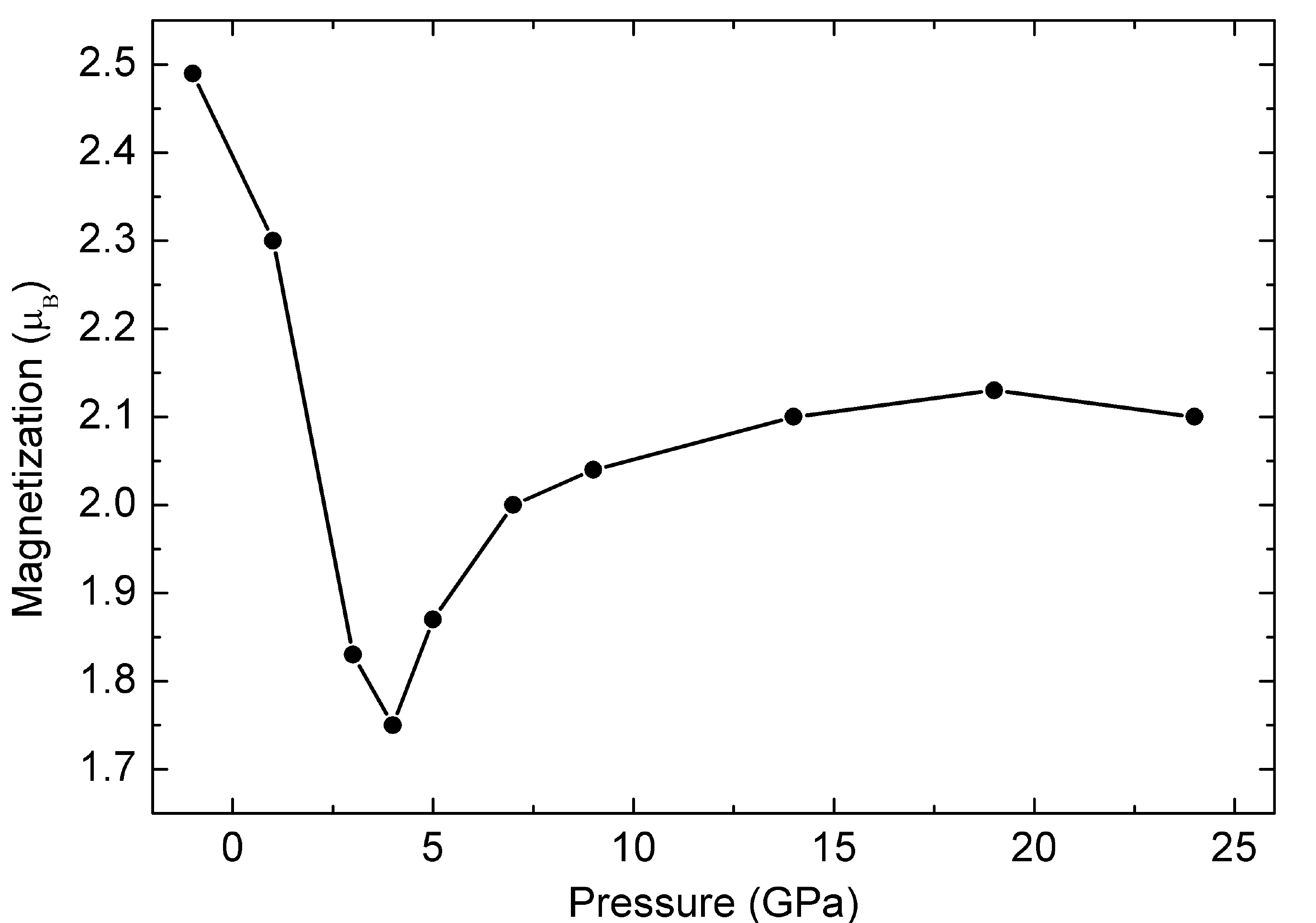}}
\caption{Calculated magnetization per unit cell of NaFe$_{2}$As$_{2}$ as a function of pressure.}
\end{figure}

\section{Summary}
From our data we find that the tetrahedral angles change continuously with pressure (inset Fig. 7) in a manner very similar to that determined previously in CaFe$_2$As$_2$ \cite{Kreyssig2008,Mittal2011} and also predicted by theory \cite{Yildirim2009}.     During the initial compression the a-axis expands and the $c$-axis rapidly decreases due to the development of the As-As interlayer bonding and this keeps Fe-As distances almost constant. In addition the tetrahedral angles are close to,  or even approach, the ideal value.   Above P$_c$ the tetrahedral angles   take value away from the ideal  because $a$-axis stops to increase to mediate the $c$-axis decrease. So, the correlation between the deviation of tetrahedral angles and $T_c$ turns out to be a \textquotedblleft side effect\textquotedblright  of the change of As atom position with increasing pressure which primary affect $T_c$ .
In conclusion, our study reveals, for the first time, the strong experimental evidence supporting the  theoretically suggested mechanism controlling the structural behavior of AFe$_{2}$As$_{2}$ compounds under pressure. Moreover the concomitant experimental and theoretical study of NaFe$_{2}$As$_{2}$ provides, missing up to now, a direct link between structural characteristics and superconductivity  under pressure.

\acknowledgments
This work was supported by the DARPA (Grants No. W31P4Q1310005 and No. W31P4Q1210008), Carnegie Canada, EFree- the DOE EFRC center at Carnegie, the Government of the Russian Federation (No. 14.A12.31.0003) and the Ministry of Education and Science of Russian Federation (Project No. 8512). GSECARS is supported by the U.S. NSF (EAR-0622171, DMR-1231586) and DOE Geosciences (DE-FG02-94ER14466). Use of the APS was supported by the DOE-BES under Contract No. DE-AC02-06CH11357. Calculations were performed on XSEDE facilities and on the cluster of the Center for Functional Nanomaterials, BNL, which is supported by the DOE-BES under contract no. DE-AC02-98CH10086. Sample growth was supported by the Natural Science Foundation of China, the \textquotedblleft Strategic Priority Research Program (B)\textquotedblright of the Chinese Academy of Sciences, and the National Basic Research Program of China.

\end{document}